# Imaging local discharge cascades for correlated electrons in WS$_2$/WSe$_2$ moiré superlattices

*Hongyuan Li[1, 2, 3, 8], Shaowei Li[1, 3, 4, 8]\*, Mit H. Naik[1, 3], Jingxu Xie[1], Xinyu Li[1], Emma Regan[1, 2, 3], Danqing Wang[1, 2], Wenyu Zhao[1], Kentaro Yumigeta[5], Mark Blei[5], Takashi Taniguchi[6], Kenji Watanabe[7], Sefaattin Tongay[5], Alex Zettl[1, 3, 4], Steven G. Louie[1, 3], Michael F. Crommie[1, 3, 4]\*, Feng Wang[1, 3, 4]\**

[1]Department of Physics, University of California at Berkeley, Berkeley, CA, USA

[2]Graduate Group in Applied Science and Technology, University of California at Berkeley, Berkeley, CA, USA

[3]Materials Sciences Division, Lawrence Berkeley National Laboratory, Berkeley, CA, USA.

[4]Kavli Energy Nano Sciences Institute at the University of California Berkeley and the Lawrence Berkeley National Laboratory, Berkeley, CA, USA

[5]School for Engineering of Matter, Transport and Energy, Arizona State University, Tempe, AZ, USA

[6]International Center for Materials Nanoarchitectonics, National Institute for Materials Science, Tsukuba, Japan

[7]Research Center for Functional Materials, National Institute for Materials Science, Tsukuba, Japan

[8]These authors contributed equally: Hongyuan Li and Shaowei Li

**Abstract:**

Transition metal dichalcogenide (TMD) moiré heterostructures provide an ideal platform to explore the extended Hubbard model[1] where long-range Coulomb interactions play a critical role in determining strongly correlated electron states. This has led to experimental observations of Mott insulator states at half filling[2-4] as well as a variety of extended Wigner crystal states at different fractional fillings[5-9]. Microscopic understanding of these emerging quantum phases, however, is still lacking. Here we describe a novel scanning tunneling microscopy (STM) technique for local sensing and manipulation of correlated electrons in a gated $WS_2/WSe_2$ moiré superlattice that enables experimental extraction of fundamental extended Hubbard model parameters. We demonstrate that the charge state of local moiré sites can be imaged by their influence on STM tunneling current, analogous to the charge-sensing mechanism in a single-electron transistor. In addition to imaging, we are also able to manipulate the charge state of correlated electrons. Discharge cascades of correlated electrons in the moiré superlattice are locally induced by ramping the STM bias, thus enabling the nearest-neighbor Coulomb interaction ($U_{NN}$) to be estimated. 2D mapping of the moiré electron charge states also enables us to determine onsite energy fluctuations at different moiré sites. Our technique should be broadly applicable to many semiconductor moiré systems, offering a powerful new tool for microscopic characterization and control of strongly correlated states in moiré superlattices.

Transition metal dichalcogenide (TMD) moiré heterostructures provide a new platform for exploring strongly correlated physics in the extended Hubbard model[1-9]. Compared with graphene-based moiré heterostructures, semiconductor TMD heterostructures feature stronger long-range Coulomb interactions and flatter moiré minibands, thus leading to new emergent correlated electronic states that are absent in graphene-based moiré systems[10-15]. Indeed, several recent studies have revealed exciting correlated states in TMD moiré superlattices, including charge-transfer insulator states[16, 17] and a rich variety of extended Wigner crystal states at fractional fillings[5-8]. Direct local characterization of the correlated states and physical parameters of the extended Hubbard model in TMD moiré heterostructures, however, has so far been lacking. For example, little is known about the strength of nearest-neighbor electron-electron interactions or the magnitude of inhomogeneity in onsite energies within TMD moiré superlattices.

Here we describe a new STM-based technique for imaging and manipulating the charge states of correlated electrons in gated $WS_2/WSe_2$ moiré superlattices that enables the determination of the nearest-neighbor Coulomb interaction energies and onsite energy fluctuations. Using this we are able to image the charge state of moiré sites via their influence on the tunneling current between an STM tip and the $WS_2/WSe_2$ heterostructure. This is analogous to Coulomb blockade in single-electron transistors where the presence of an additional electron can dramatically modulate electrical transport. By combining a back-gate voltage with the STM bias this mechanism enables us to locally charge and discharge correlated moiré electrons. Gradually ramping the STM bias under these conditions results in a cascade of discharging events for correlated electrons at multiple neighboring moiré sites. Systematic investigation of this discharge cascade allows determination of nearest-neighbor Coulomb interactions as well as onsite energy fluctuations within the moiré superlattice.

A schematic diagram of our aligned WSe$_2$/WS$_2$ heterostructure device is shown in Fig. 1a. We used an array of graphene nanoribbons (GNRs) with 100~200 nm separation as a top contact electrode, and the doped silicon substrate as a back-gate to control the global carrier density within the heterostructure. This configuration has proven to work reliably for STM study of TMD materials on insulating substrates down to liquid helium temperatures[18]. Details of the device fabrication are presented in the reference[18]. Figure 1b shows an ultra-high vacuum (UHV) STM image of the moiré superlattice in an exposed WSe$_2$/WS$_2$ area between two graphene nanoribbons at T= 5.4K. Three types of moiré sites are labeled: AA, B$^{Se/W}$, and B$^{W/S}$, with the corresponding chemical structures shown in the top-view sketch of Fig. 1c. The moiré period is $l_m = 8.1$nm, indicating a near-zero twist angle.

We characterized our gated WSe$_2$/WS$_2$ heterostructure via scanning tunneling spectroscopy (STS). Figs. 1d and 1f show plots of the differential conductivity (dI/dV) at the B$^{Se/W}$ and B$^{W/S}$ sites respectively as a function of both the STM tip bias V$_b$ and backgate voltage V$_g$. Here we focus on the electron-doped regime of the moiré heterostructure at positive V$_g$, with the Fermi level located in the conduction band. The dI/dV spectra show negligible signal for 0 < V$_b$ < 0.4V due to very small tunneling probability to conduction band-edge states which lie at the K point of the bottom WS$_2$ layer and which feature a large out-of-plane decay constant[18]. At V$_b$ > 0.4 V tip electrons can more readily tunnel into the conduction band Q-point states which have a smaller decay constant (i.e., they protrude further into the vacuum).

In addition to a general increase in dI/dV signal for V$_b$ > 0.4 V, we observe sharp dispersive dI/dV peaks at the different moiré sites (the bright features labeled by blue arrows in Figs. 1d, f). These dispersive peaks can be more clearly observed in the density plot of the second order derivative ($d^2I/dV^2$), as shown in Fig. 1e,g. To better understand the origin of these peaks we performed 2D dI/dV mapping at STM bias voltages of $V_b$ = 0.775 V (Fig. 1h) and $V_b$ = 0.982 V (Fig. 1i) for a fixed gate voltage of $V_g$=45V. In these images, the dispersive dI/dV peaks in Figs. 1d, f correspond to periodic circular rings surrounding the $B^{Se/W}$ sites that expand upon increased $V_b$. This behavior can be explained by discrete charging/discharging events for localized moiré electrons at the $B^{Se/W}$ sites, controlled by the application of $V_g$ and $V_b$. The gain or loss of an electron at a $B^{Se/W}$ moiré site dramatically modifies the electron tunneling rate between the STM tip and the heterostructure surface, analogous to how localized electrons modulate electrical conductance through single-electron transistors via Coulomb blockade effects[19]. Charging/discharging events at the $B^{Se/W}$ sites thus lead to a corresponding jump in the STM tunneling current and result in sharp peaks in the dI/dV spectra (similar ring-like charging behavior has been seen via STM in other nanoscale systems[20-24]).

*Ab initio* calculations show that the $WSe_2/WS_2$ moiré flat band states at the conduction band edge are strongly localized at the $B^{Se/W}$ sites in real space (see Fig. S1 in the SI). For $V_g$ > 25V, the moiré global filling factor $n/n_0$ is greater than 1, where n is the gate-controlled carrier density and $n_0$ is the carrier density corresponding to a half-filled moire miniband with one electron per moire lattice site. At this gate-voltage there is thus at least one electron localized at each $B^{Se/W}$ site. The effect of the STM tip is that it behaves as a local top gate that modifies nearby moiré-site electron energy. At positive sample bias, $V_b$, negative charge accumulates at the tip and repels nearby electrons, thus causing electrons localized to $B^{Se/W}$ sites to discharge when $V_b$ exceeds a threshold value (see the sketch in Fig. 1j). The efficiency with which the STM tip discharges nearby localized electrons depends sensitively on their distance to the tip, and thus results in circular discharging rings for a given $V_b$ (Fig. 1h). These rings expand continuously with increased $V_b$ since larger $V_b$ enables the discharge of localized electrons at larger tip-electron distances (Fig. 1i). When the tip is inside a discharge ring the circled moiré site is empty whereas it contains an electron when the tip is outside the ring.

The STM tip can discharge multiple correlated electrons at neighboring moiré sites in a cascade fashion with increased $V_b$, thus providing a powerful tool to probe electron correlation in moiré systems. We have systematically examined the discharge cascade of correlated electrons in a $WSe_2/WS_2$ flat-band by performing 2D dI/dV mapping of the moiré pattern as a function of $V_b$. Figs. 2a-i show the evolution of discharge rings as $V_b$ is increased from 0.57 V to 1.25 V for fixed $V_g$ = 52 V. The discharge rings expand with increased $V_b$ and begin crossing each other at $V_b$ = 0.66 V. Near a crossing point, the STM tip couples effectively to multiple adjacent moiré sites and can generate a cascade of discharging events. Intricate new patterns emerge as the rings cross each other with increasing $V_b$. These patterns differ distinctly from a simple superposition of ever-increasing rings as expected from a non-interacting picture, thus providing a manifestation of electron correlation in TMD moiré superlattices.

The effects of electron correlation on cascade discharging can be more effectively visualized in position-dependent dI/dV spectra. Fig. 3a shows position-dependent dI/dV spectra along the green line in Fig. 2f. This line passes through a high-symmetry 2-ring crossing point (marked D) which is equidistant to neighboring $B^{Se/W}$ sites I and II. For positions near D sites I and II are both occupied at low $V_b$ (i.e., $n = 2$ where $n$ is the total electron count for adjacent moiré sites). As $V_b$ increases, however, the tip successively discharges the two sites and $n$ changes from 2 to 1 and then from 1 to 0 as $V_b$ crosses two dI/dV discharge peaks. At D one would expect by symmetry that these two discharging events should occur at the same value of $V_b$ for a non-interacting picture. The data of Fig. 3a, however, show that these two discharging events occur at different $V_b$ values, with a discharging gap of $\Delta V_b^D = 122 \pm 9$ mV (obtained via high-resolution dI/dV mapping, see SI).

Similar behavior can be seen at another high symmetry point, marked T in Fig. 2f. Here the STM tip is equidistant to three neighboring $B^{Se/W}$ sites marked I, II, and III, and so the discharge cascade involves three electrons. Fig. 3e shows position-dependent dI/dV spectra along the yellow line in Fig. 2f which passes through T. Three dI/dV peaks are seen in the spectra, corresponding to a cascade of three discharging events that decrease the total number of electrons (*n*) in sites I-III from 3 to 2 to 1 and then, finally, to 0. At T we observe the voltage difference between the 3→2 and 2→1 discharge peaks to be identical to the difference between the 2→1 and 1→0 discharge peaks within the uncertainty of our measurement: $\Delta V_b^T = 166$ mV $\pm$ 11 mV.

In order to interpret the discharge cascade of correlated electrons in our TMD moiré system in terms of physically significant parameters, we employ a simplified *N*-moiré-site model that includes on-site and nearest-neighbor interactions. The Hamiltonian describing our system is

$$H = \sum_{i=1}^{N}(\varepsilon_i + v_i)n_i + \frac{1}{2}U_{NN}\sum_{<ij>}n_i n_j. \quad (1)$$

Here $U_{NN}$ is the nearest-neighbor Coulomb interaction term and $<ij>$ only sums over nearest neighbors and *N* represents the number of moiré sites close to the tip (equals to 2 or 3 in our system). $n_i$ is the electron number at moiré site *i*, $\varepsilon_i$ is the onsite energy at site *i*, and $v_i$ is the potential energy shift induced by $V_b$ and $V_g$ at site *i*. $v_i$ has the form

$$v_i = \alpha_i \cdot eV_b - \beta_i \cdot eV_g \quad (2)$$

where $e = 1.6 \times 10^{-19} C$, and $\alpha_i$ and $\beta_i$ are dimensionless coefficients describing the electrostatic potential on site $i$ induced by V$_b$ and V$_g$, respectively. We note that $\alpha_i$ depends sensitively on the tip position $r_t$, meaning $\alpha_i = \alpha_i(r_t)$. In this model we neglect intersite hopping due to the small bandwidth of the moiré flat band (~5meV, see SI). We also ignore on-site Coulomb interactions since the total number of electrons ($n = \sum_i n_i$) for sites near the tip is smaller than $N$ and the energy of double occupancy for a single site is assumed to be prohibitively large. This Hamiltonian is meant to describe electrons in the lowest conduction band near the tip since higher-energy delocalized electrons are assumed to be swept away by tip repulsion for V$_b$ > 0.

Our strategy for understanding the discharge phenomena observed here is to explore the consequences of this Hamiltonian for different electron occupation values $n$. By comparing the different total energies, E($n$), we can identify the charge occupation, $n$*, that has the lowest total energy and we assume that this is the ground state. A discharge event from the $n = n$* + 1 state to the $n = n$* state occurs when E($n$*) < E($n$* + 1). Since our measurements primarily involve discharging events, the largest relevant energy is $U_{NN}$ which eclipses the energy associated with intersite hopping. As a result, the behavior induced by (1) can be treated within an essentially classical framework (in the context of discharge phenomena) that is adequate to extract information on the Hubbard model parameters $\varepsilon_i$ and $U_{NN}$ from our data, the main goal of this work.

We start by applying this model to analyze the discharge behavior that occurs when the STM tip is held at position D. Here the tip is equidistant from sites I and II (the closest moiré sites to the tip) and so we model the moiré system as an N = 2 cluster as illustrated in Fig. 3b. The on-site energy (Eq. (1)) for sites I and II can be written as $\varepsilon$ and the electrostatic potential energy (Eq. (2)) for each site is $v = e\alpha(r^D)V_b - e\beta V_g$ where $r^D = 3.9$ nm is equidistant from sites I and II. Straightforward energetic considerations allow the N=2 ground state energy for different $n$ to be written as $E(n=2) = 2(\varepsilon + v) + U_{NN}$, $E(n=1) = \varepsilon + v$, and $E(n=0) = 0$. Fig. 3c shows a plot of $E(2)$, $E(1)$, and $E(0)$ as a function of applied electrostatic potential, $v$. Three different regimes can be seen where the ground state energy transitions from an $n = 2$ charge state to an $n = 1$ charge state and then to an $n = 0$ charge state as $v$ is increased. The boundaries between these different charge states mark the locations of electron discharging events. The first discharging event happens when $E(2) = E(1)$, which occurs at the potential $v^1 = -U_{NN} - \varepsilon$. The second discharging event happens when $E(1) = E(0)$, which occurs at the potential $v^2 = -\varepsilon$. The difference in electrostatic potential energy between these two discharge events is then $\Delta v = v^2 - v^1 = U_{NN}$. Using Eq. (2) and assuming that the gate voltage remains unchanged while ramping the bias voltage (typical for our experiments) allows $U_{NN}$ to be expressed in terms of the first and second discharge bias voltages: $U_{NN} = e\alpha(r^D)(V_2 - V_1) = e\alpha(r^D)\Delta V^D$.

The discharge cascade behavior when the tip is located at T can be analyzed using similar reasoning, except for an N=3 cluster instead of an N=2 cluster. In this case $v = e\alpha(r^T)V_b - e\beta V_g$, where $r^T = 4.7 nm$ is the distance between T and the three neighboring $B^{Se/W}$ sites. The resulting energy levels for different $n$ are found to be $E(3) = 3(\varepsilon + v) + 3U_{NN}$, $E(2) = 2(\varepsilon + v) + U_{NN}$, $E(1) = \varepsilon + v$, and $E(0) = 0$, which are plotted in Fig. 3g. Discharging events in the 3-site moiré system thus occur at $v_1 = -\varepsilon - 2U_{NN}$, $v_2 = -\varepsilon - U_{NN}$, and $v_3 = -\varepsilon - U_{NN}$ (Fig. 3h). This provides an additional means of finding the nearest-neighbor Coulomb interaction energy $U_{NN} = v_3 - v_2 = v_2 - v_1 = \alpha(r^T)e\Delta V_b^T$ by utilizing the voltage difference, $\Delta V_b^T$, between discharge events at T.

A similar analysis can be used to determine variations in the Hubbard on-site energy, $\varepsilon_i$, for a moiré superlattice. This comes from the fact that for small $V_b$ the STM tip can only discharge a single moiré site whose energy is described by $E(1) = \varepsilon + v$ and $E(0) = 0$ (i.e., the N=1 limit). In this case, discharge happens when $E(1) = E(0)$, which occurs when $v = -\varepsilon$. Fluctuations in $\varepsilon$ are thus directly related to fluctuations in the discharge potential, $\delta\varepsilon = -\delta v$, which (using Eq. (2)) leads to $\delta\varepsilon = -e\alpha(r_t)\delta V_b$ where $\delta V_b$ represents spatial fluctuations in the measured single-site discharge voltage.

This type of behavior can be seen experimentally as shown in Fig. 4. Fig. 4a shows a dI/dV map of a pristine region of the WSe$_2$/WS$_2$ moiré superlattice for $V_g$ = 50 V and $V_b$ = 0.465 V. The discharge rings around the B$^{Se/W}$ moiré sites are quite uniform in this defect-free region. This uniformity is also seen in a dI/dV spectra linecut (Fig. 4b) that goes through five moiré sites along the red line in Fig. 4a. Fig. 4c, on the other hand, shows a dI/dV map obtained near a point defect (marked by a red dot) for a similar set of parameters ($V_g$ = 53 V and $V_b$ = 0.740 V). Here the discharge rings are highly non-uniform (the defect moiré site itself does not show a clear discharge ring for this set of parameters due to the large change in its onsite energy). The magnitude of the effect of the defect on neighboring moiré sites can be seen through dI/dV spectra (Fig. 4d) obtained along the red linecut shown in Fig. 4c. As seen in Fig. 4d, the defect causes significant changes in the onsite energies of adjacent moiré sites. The discharge bias (measured at the discharge ring center, $r_t = 0$) of sites I and II, for example, is ~200 mV lower than those for sites III and IV (see blue dashed line in Fig. 4d). This implies that the on-site energy shift on sites I and II is $\delta\varepsilon \approx \alpha(0)(200 meV)$.

A problem with our characterization of moiré Hubbard parameters up to now is that we cannot convert them to quantitative energies until we determine $\alpha(r_t)$, the geometric electrostatic conversion factor of Eq. (2). In particular, we require $\alpha(r^D)$, $\alpha(r^T)$, and $\alpha(0)$ to quantitatively determine $U_{NN}$ and $\delta\varepsilon$. We can gain some experimental insight into the behavior of $\alpha(r_t)$ from the slopes of the lines representing discharge peaks in the dI/dV plots of Figs.1d and 1e. The condition for discharge in these plots is $v = \alpha(r_t) \cdot eV_b - \beta \cdot eV_g = -C$, where $C$ is a constant independent of $V_b$ and $V_g$. This can be rewritten as $V_g = \frac{\alpha(r_t)}{\beta} V_b + \frac{C}{e\beta}$, which defines the linear discharge traces observed in Figs. 1d and 1e. The experimental slope of the discharge traces thus yields the ratio $\frac{\alpha(r_t)}{\beta}$. This does not provide the precise value of $\alpha(r_t)$ (since the back-gate factor $\beta$ is still unknown), but by assuming that $\beta$ is constant we can experimentally determine the proportionality of $\alpha(r_t)$ for different tip locations from the slopes of the discharge traces at those locations. We have done this for points D, T, and $B^{Se/W}$, thereby enabling us to determine the ratios $\alpha(r^D) : \alpha(r^T) : \alpha(0) = 0.53 : 0.39 : 1$ (see SI).

Obtaining a quantitative value of $\alpha(r_t)$, however, remains difficult since the magnitudes of $\alpha(r_t)$ and $\beta$ depend on non-universal details of the experimental setup. They involve, for example, capacitive coupling between the tip, back gate, and different layers of our device heterostructure. To overcome this problem we utilized the COMSOL software package to numerically solve Poisson's equation for our specific device geometry (see SI). Modeling our STM tip as a metallic cone with a cone-angle of 30° and a tip-surface height of 0.8 nm results in a ratio of $\alpha(r^D) : \alpha(r^T) : \alpha(0) = 0.55 : 0.49 : 1$, in reasonable agreement with the experimental ratios above. A significant benefit of the numerical simulation is that it provides the absolute magnitude of the $\alpha(r_t)$ factors: $\alpha(r^D) = 0.18$, $\alpha(r^T) = 0.16$, and $\alpha(0) = 0.33$. These factors allow us to extract a quantitative value of $U_{NN} = 22 \pm 2$ meV from our measurements at D and $U_{NN} = 27 \pm 2$ meV from our measurements at T. These two values of $U_{NN}$ are in reasonable agreement with each other, a self-consistency check that helps to validate our overall approach. We are also able to determine the fluctuation in onsite energy of sites I and II around the point defect in Fig. 4c to be $\delta\varepsilon \sim 65\ meV$.

The expected value of $U_{NN}$ can be roughly estimated by considering the energy difference associated with the initial position ($\vec{r}_i$) of the discharging electron and its final position ($\vec{r}_f$) after discharge. For an electron being discharged from the N=2 cluster discussed above for point D, the initial electrostatic energy is $E(\vec{r}_i) \approx \frac{1}{4\pi\epsilon_{eff}\epsilon_0} \cdot \frac{e^2}{l_m}$ (assuming each moire site contains one electron). In order to discharge, an electron need only escape a screening distance away from the STM tip. Although the screening distance is difficult to determine accurately, since only two lattice sites participate in discharge events at D (Fig. 3a) we can estimate it to be on the order of $l_m$. An electron thus needs only to hop a distance ~$l_m$ to escape, thus placing the electron in some final configuration with a residual Coulomb energy of the order of $\frac{e^2}{\eta \cdot l_m}$, where $\eta \cdot l_m$ is the distance of the escaped electron from the remaining electron. Taking $\eta$ to be approximately 2, the electron's energy difference is then $U_{NN} = E(\vec{r}_i) - E(\vec{r}_f) \approx \frac{1}{4\pi\epsilon_{eff}\epsilon_0} \cdot \left(\frac{e^2}{l_m} - \frac{e^2}{2l_m}\right) = \frac{1}{4\pi\epsilon_{eff}\epsilon_0} \cdot \frac{e^2}{2l_m}$. Here the effective dielectric constant is $\epsilon_{eff} = \frac{1}{2}(\epsilon_{vac} + \epsilon_{hBN})$ where $\epsilon_{vac} = 1$ and $\epsilon_{hBN}$ is the dielectric constant of hBN. Since the dielectric constant of hBN is anisotropic, we approximate it as $\epsilon_{hBN} = \frac{1}{2}(\epsilon_\perp + \epsilon_\parallel)$ where $\epsilon_\perp = 4$ and $\epsilon_\parallel = 7$ are the out-of-plane and in-plane hBN dielectric constants, respectively[25].

Taken together, these parameters yield an expected value of $U_{NN} \approx 30\text{meV}$, reasonably consistent with our experimental value of $U_{NN} \approx 25\text{meV}$ obtained by averaging the $U_{NN}$ values measured at D and T. We note that this value of $U_{NN}$ is significantly larger than the bandwidth $W$

≈ 5meV obtained from *ab initio* calculations of the conduction miniband for $WSe_2/WS_2$ moiré heterostructures (see SI). This confirms that $WSe_2/WS_2$ heterostructures lie in the strongly correlated limit, consistent with previous observations of the extended Wigner crystal states in TMD moiré superlattices[5-8].

In conclusion, we have demonstrated a new technique for imaging and manipulating the local charge state of correlated electrons in gated $WS_2/WSe_2$ moiré superlattices using the tip of an STM. Observation of a cascade of local electron discharging events enables us to obtain key microscopic parameters of the extended Hubbard model that describes this moiré system. Our technique should be broadly applicable to other types of moiré systems, offering a powerful new tool for microscopic characterization and control of strongly correlated states in moiré superlattices.

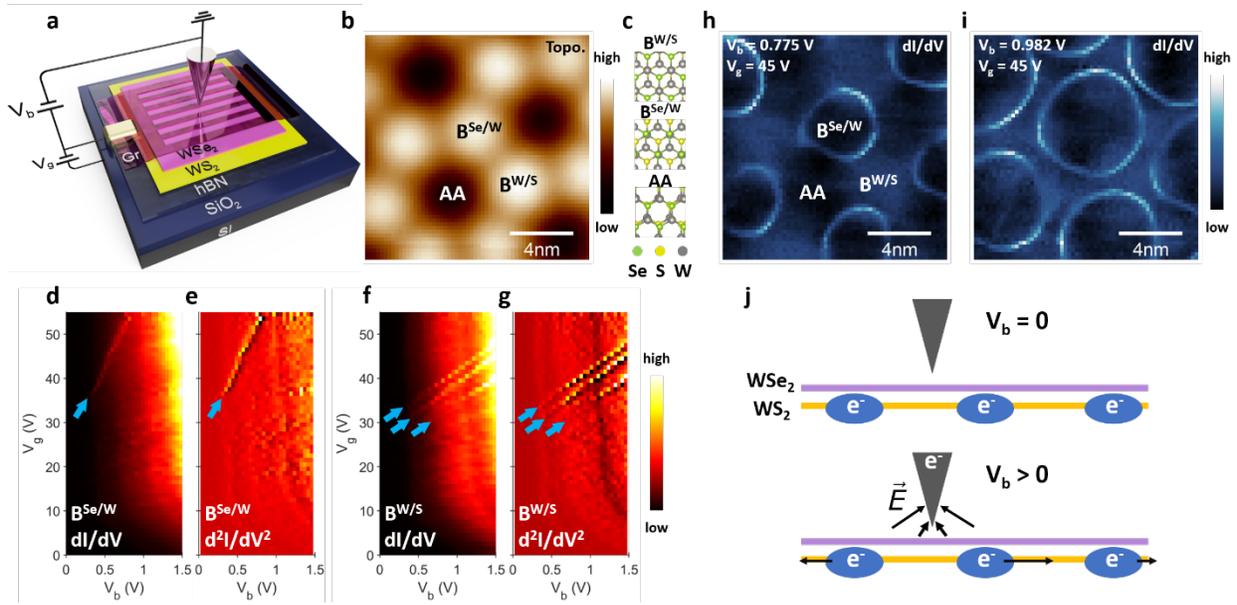

**Figure 1. Local discharging of electrons at moiré sites by an STM tip. a**. Schematic of the gate-tunable WSe$_2$/WS$_2$ heterostructure device used for this STM study. Graphene nanoribbons (Gr) are placed on top of the WSe$_2$/WS$_2$ as contact electrodes. **b**. UHV STM image of an exposed WSe$_2$/WS$_2$ area (T = 5.4K). A moiré superlattice is clearly resolved. $V_b$ = -3V, I = 100pA, $V_g$ = 45 V. **c**. Schematic of the three high-symmetry stacking configurations: $B^{W/S}$, $B^{Se/W}$, and AA. **d-g**. Gate-dependent dI/dV spectra measured at the (**d**) $B^{Se/W}$ site and (**f**) $B^{W/S}$ site. The bright dispersive features labeled by blue arrows correspond to tip-induced charging/discharging of electrons at adjacent $B^{Se/W}$ moiré sites. To highlight the dispersive feature, the corresponding second order derivative ($d^2I/dV^2$) spectra are plotted in **e** and **g**, respectively. **h,i**. dI/dV maps of the region shown in **b** measured with (**h**) $V_b$ = 0.775 V, $V_g$ = 45 V and (**i**) $V_b$ = 0.982 V, $V_g$ = 45 V. The tip-sample distance for the dI/dV mappings is controlled by the setpoint: $V_b$ = -3V, I = 100 pA. Discharge of the moiré electron causes a dI/dV peak which appears as a circular ring surrounding each $B^{Se/W}$ site in the moiré superlattice. The periodic discharging rings expand with $V_b$. **j**. Sketch of tip-induced moiré electron discharging. The STM tip acts as a local top gate that modifies the electron energy at nearby $B^{Se/W}$ moiré sties and discharges them for bias voltages greater than a local threshold value.

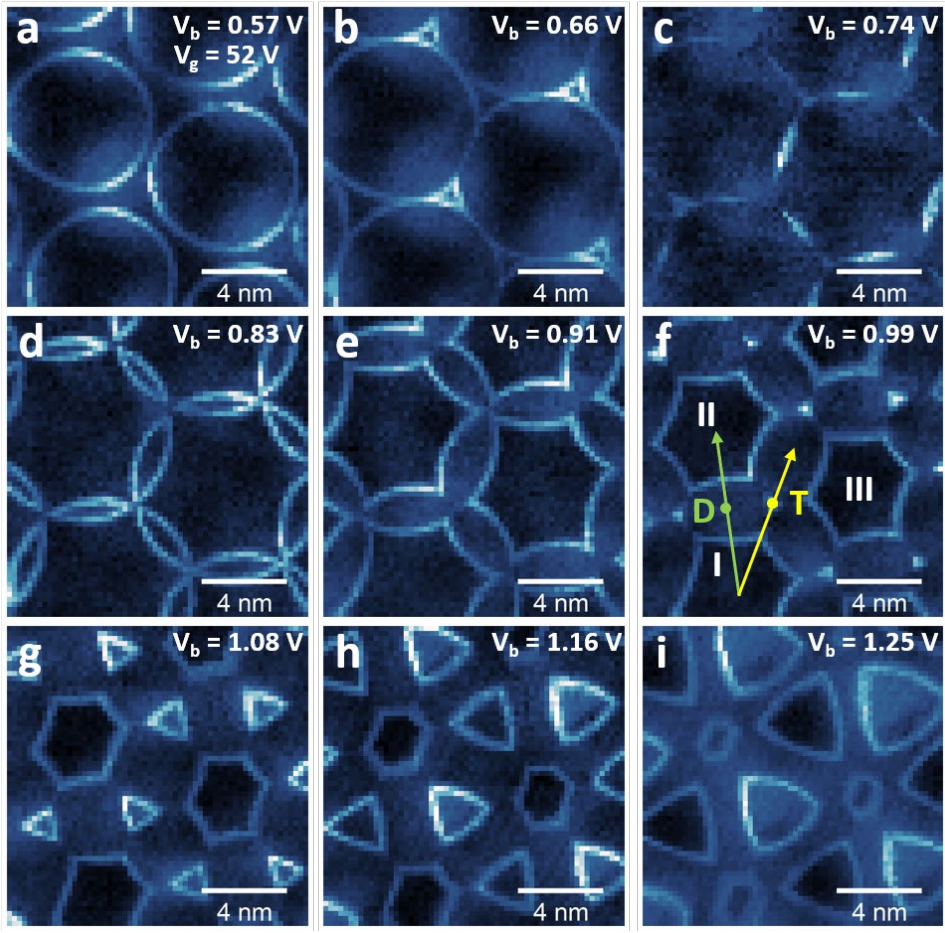

**Figure 2. Discharging correlated electrons. a-i**. Evolution of moiré discharging rings with increasing sample bias $V_b$ (gate voltage fixed at $V_g = 52$ V). The rings expand with increased $V_b$ and intricate new patterns emerge when they cross each other. The complex patterns indicate discharging cascades of multiple correlated electrons in neighboring moiré sites. dI/dV spectra along the green and yellow lines in **f** are shown in Fig. 3. Here D and T denote the 2-ring and 3-ring crossing points, respectively. dI/dV mapping setpoint: $V_b = -3$V, I = 100 pA, $V_g = 52$ V. (T = 5.4 K).

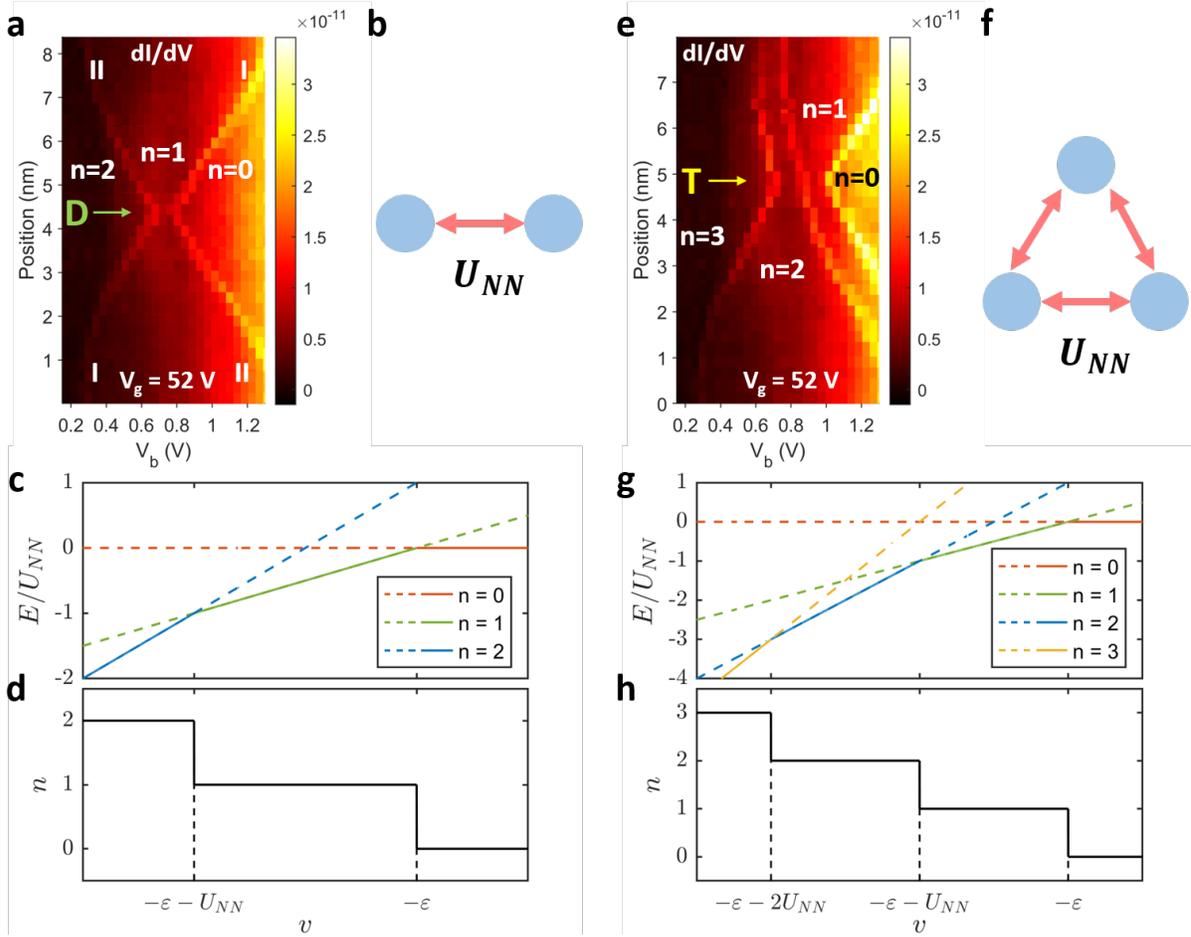

**Figure 3. Correlation effects on cascade discharging of moiré sites. a**. Position-dependent dI/dV spectra along the green line shown in Fig. 2f which passes through D. dI/dV peaks indicated by bright lines correspond to discharging events where the total electron number decreases from 2 to 1 and from 1 to 0 from left to right. I and II indicate the moiré sites being discharged as labeled in Fig. 2f. **b.** Sketch of simplified 2-site cluster model for analysis of discharge behavior at D. $U_{NN}$ indicates the nearest-neighbor Coulomb interaction. **c**. Calculated energy levels and (**d**) total electron number ($n$) of cluster ground state as a function of $v$ for the 2-site model. States with different electron number are labeled by color. The discrete changes of $n$ in **d** correspond to discharge peaks at B. **e.** Position-dependent dI/dV spectra along the yellow line shown in Fig. 2f. This line cut passes through T. Bright lines indicate discharging events at three distinct bias voltages. **f**. Sketch of simplified 3-site cluster model for analysis of discharge behavior at T. **g**. Calculated energy levels and (**h**) total electron number of cluster ground state as a function of $v$ for the 3-site model.

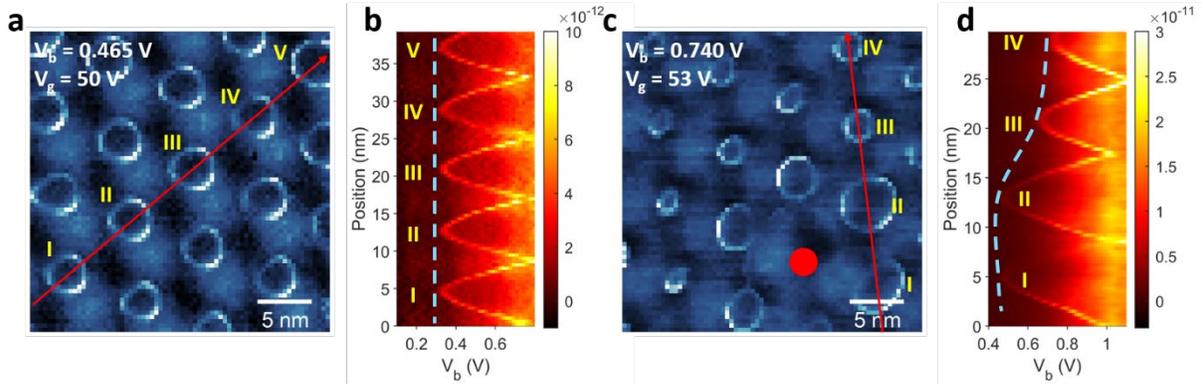

**Figure 4. Inhomogeneity of moiré onsite energy. a.** dI/dV map of a representative homogeneous region of the WS$_2$/WSe$_2$ moiré superlattice ($V_b$ = 0.465 V, $V_g$ = 50 V). **b**. dI/dV spectra measured along the red linecut shown in **a**. Discharge voltages at moiré sites I - V are seen to be nearly uniform. **c**. dI/dV map of discharge rings close to a point defect shows strongly non-uniform behavior. Solid red circle marks the position of the point defect ($V_b$ = 0.740 V, $V_g$ = 53 V). **d.** dI/dV spectra measured along the red linecut shown in **c**. A significant reduction in discharge bias is observed for sites I and II near the defect. Tip-sample distance is determined by the setpoint: $V_b$ = -3V, I = 100 pA.

**Corresponding Author**

* Email: swli@berkeley.edu (S.L.), crommie@physics.berkeley.edu (M.C.) and fengwang76@berkeley.edu (F.W.)

**Author Contributions**

F.W. and M.C. conceived the project. S.G.L. supervised the theoretical calculations. H.L., S.L. performed the STM measurement, M.H.N. carried out the DFT calculations. H.L., J.X., X.L., W.Z., E.R., and D.W. fabricated the heterostructure device. performed the SHG measurement. K.Y., M.B. and S.T. grew $WSe_2$ and $WS_2$ crystals. K.W. and T.T. grew the hBN single crystal. All authors discussed the results and wrote the manuscript.

**Notes**

The authors declare no financial competing interests.

**ACKNOWLEDGMENT**

This work was funded by the U.S. Department of Energy, Office of Science, Office of Basic Energy Sciences, Materials Sciences and Engineering Division under Contract No. DE-AC02-05-CH11231 (van der Waals heterostructure program KCFW16) (device electrode preparation, STM spectroscopy, DFT calculations and theoretical analysis). Support was also provided by the US Army Research Office under MURI award W911NF-17-1-0312 (device layer transfer), by the National Science Foundation Award DMR-1926004 (structural determination), and by the National Science Foundation Award DMR-1807233 (surface preparation). S.T.

acknowledges support from DOE-SC0020653, NSF 1935994, NSF DMR 1552220, DMR 1904716, 1955889, NSF CMMI 1933214, and Applied Materials Inc. for defects engineering, crystal growth technologies, and crystalline analysis towards $WSe_2$ and $WS_2$ bulk crystal development. K.W. and T.T. acknowledge support from the Elemental Strategy Initiative conducted by the MEXT, Japan, Grant Number JPMXP0112101001, JSPS KAKENHI Grant Number JP20H00354 and the CREST(JPMJCR15F3), JST for bulk hBN crystal growth and analysis. E.C.R. acknowledges support from the Department of Defense (DoD) through the National Defense Science & Engineering Graduate Fellowship (NDSEG) Program. S.L. acknowledges support from Kavli ENSI Heising Simons Junior Fellowship. M.H.N. thanks Sudipta Kundu and Manish Jain for their implementation of noncollinear wavefunction plotting in Siesta.

**Supplementary Materials**

**Data availability**

The data supporting the findings of this study are included in the main text and in the Supplementary Information files, and are also available from the corresponding authors upon reasonable request.

**Reference**

1. Wu, F.; Lovorn, T.; Tutuc, E.; MacDonald, A. H. *Hubbard model physics in transition metal dichalcogenide moiré bands. Physical review letters* **2018,** 121, (2), 026402.
2. Wang, L.; Shih, E.-M.; Ghiotto, A.; Xian, L.; Rhodes, D. A.; Tan, C.; Claassen, M.; Kennes, D. M.; Bai, Y.; Kim, B. *Correlated electronic phases in twisted bilayer transition metal dichalcogenides. Nature materials* **2020**, 1-6.
3. Tang, Y.; Li, L.; Li, T.; Xu, Y.; Liu, S.; Barmak, K.; Watanabe, K.; Taniguchi, T.; MacDonald, A. H.; Shan, J. *Simulation of Hubbard model physics in WSe 2/WS 2 moiré superlattices. Nature* **2020,** 579, (7799), 353-358.
4. Shimazaki, Y.; Schwartz, I.; Watanabe, K.; Taniguchi, T.; Kroner, M.; Imamoğlu, A. *Strongly correlated electrons and hybrid excitons in a moiré heterostructure. Nature* **2020,** 580, (7804), 472-477.

5. Regan, E. C.; Wang, D.; Jin, C.; Utama, M. I. B.; Gao, B.; Wei, X.; Zhao, S.; Zhao, W.; Zhang, Z.; Yumigeta, K. *Mott and generalized Wigner crystal states in WSe 2/WS 2 moiré superlattices. Nature* **2020,** 579, (7799), 359-363.
6. Jin, C.; Tao, Z.; Li, T.; Xu, Y.; Tang, Y.; Zhu, J.; Liu, S.; Watanabe, K.; Taniguchi, T.; Hone, J. C. *Stripe phases in WSe2/WS2 moir\'e superlattices. arXiv preprint arXiv:2007.12068* **2020**.
7. Xu, Y.; Liu, S.; Rhodes, D. A.; Watanabe, K.; Taniguchi, T.; Hone, J.; Elser, V.; Mak, K. F.; Shan, J. *Abundance of correlated insulating states at fractional fillings of WSe $_{2}$/WS $_{2}$ moir\'e superlattices. arXiv preprint arXiv:2007.11128* **2020**.
8. Huang, X.; Wang, T.; Miao, S.; Wang, C.; Li, Z.; Lian, Z.; Taniguchi, T.; Watanabe, K.; Okamoto, S.; Xiao, D. *Correlated Insulating States at Fractional Fillings of the WS2/WSe2 Moir\'e Lattice. arXiv preprint arXiv:2007.11155* **2020**.
9. Chu, Z.; Regan, E. C.; Ma, X.; Wang, D.; Xu, Z.; Utama, M. I. B.; Yumigeta, K.; Blei, M.; Watanabe, K.; Taniguchi, T. *Nanoscale Conductivity Imaging of Correlated Electronic States in WSe 2/WS 2 Moiré Superlattices. Physical Review Letters* **2020,** 125, (18), 186803.
10. Liu, X.; Hao, Z.; Khalaf, E.; Lee, J. Y.; Ronen, Y.; Yoo, H.; Najafabadi, D. H.; Watanabe, K.; Taniguchi, T.; Vishwanath, A. *Tunable spin-polarized correlated states in twisted double bilayer graphene. Nature* **2020,** 583, (7815), 221-225.
11. Lu, X.; Stepanov, P.; Yang, W.; Xie, M.; Aamir, M. A.; Das, I.; Urgell, C.; Watanabe, K.; Taniguchi, T.; Zhang, G. *Superconductors, orbital magnets and correlated states in magic-angle bilayer graphene. Nature* **2019,** 574, (7780), 653-657.
12. Cao, Y.; Fatemi, V.; Demir, A.; Fang, S.; Tomarken, S. L.; Luo, J. Y.; Sanchez-Yamagishi, J. D.; Watanabe, K.; Taniguchi, T.; Kaxiras, E. *Correlated insulator behaviour at half-filling in magic-angle graphene superlattices. Nature* **2018,** 556, (7699), 80.
13. Zondiner, U.; Rozen, A.; Rodan-Legrain, D.; Cao, Y.; Queiroz, R.; Taniguchi, T.; Watanabe, K.; Oreg, Y.; von Oppen, F.; Stern, A. *Cascade of phase transitions and Dirac revivals in magic-angle graphene. Nature* **2020,** 582, (7811), 203-208.
14. Cao, Y.; Rodan-Legrain, D.; Rubies-Bigorda, O.; Park, J. M.; Watanabe, K.; Taniguchi, T.; Jarillo-Herrero, P. *Tunable correlated states and spin-polarized phases in twisted bilayer–bilayer graphene. Nature* **2020**, 1-6.
15. Chen, G.; Jiang, L.; Wu, S.; Lyu, B.; Li, H.; Chittari, B. L.; Watanabe, K.; Taniguchi, T.; Shi, Z.; Jung, J. *Evidence of a gate-tunable Mott insulator in a trilayer graphene moiré superlattice. Nature Physics* **2019,** 15, (3), 237-241.
16. Slagle, K.; Fu, L. *Charge Transfer Excitations, Pair Density Waves, and Superconductivity in Moir\'e Materials. arXiv preprint arXiv:2003.13690* **2020**.
17. Zhang, Y.; Yuan, N. F.; Fu, L. *Moir\'e quantum chemistry: charge transfer in transition metal dichalcogenide superlattices. arXiv preprint arXiv:1910.14061* **2019**.
18. Li, H.; Li, S.; Naik, M. H.; Xie, J.; Li, X.; Wang, J.; Regan, E.; Wang, D.; Zhao, W.; Zhao, S. *Imaging moir\'e flat bands in 3D reconstructed WSe2/WS2 superlattices. arXiv preprint arXiv:2007.06113* **2020**.
19. Kastner, M. A. *The single-electron transistor. Reviews of modern physics* **1992,** 64, (3), 849.
20. Jung, S.; Rutter, G. M.; Klimov, N. N.; Newell, D. B.; Calizo, I.; Hight-Walker, A. R.; Zhitenev, N. B.; Stroscio, J. A. *Evolution of microscopic localization in graphene in a magnetic field from scattering resonances to quantum dots. Nature Physics* **2011,** 7, (3), 245-251.
21. Pradhan, N. A.; Liu, N.; Silien, C.; Ho, W. *Atomic scale conductance induced by single impurity charging. Physical review letters* **2005,** 94, (7), 076801.
22. Brar, V. W.; Decker, R.; Solowan, H.-M.; Wang, Y.; Maserati, L.; Chan, K. T.; Lee, H.; Girit, Ç. O.; Zettl, A.; Louie, S. G. *Gate-controlled ionization and screening of cobalt adatoms on a graphene surface. Nature Physics* **2011,** 7, (1), 43-47.

23. Wong, D.; Velasco, J.; Ju, L.; Lee, J.; Kahn, S.; Tsai, H.-Z.; Germany, C.; Taniguchi, T.; Watanabe, K.; Zettl, A. *Characterization and manipulation of individual defects in insulating hexagonal boron nitride using scanning tunnelling microscopy. Nature nanotechnology* **2015,** 10, (11), 949-953.
24. Teichmann, K.; Wenderoth, M.; Loth, S.; Ulbrich, R.; Garleff, J.; Wijnheijmer, A.; Koenraad, P. *Controlled charge switching on a single donor with a scanning tunneling microscope. Physical review letters* **2008,** 101, (7), 076103.
25. Geick, R.; Perry, C.; Rupprecht, G. *Normal modes in hexagonal boron nitride. Physical Review* **1966,** 146, (2), 543.

# Supplementary Information for
# Imaging local discharge cascades for correlated electrons in WS$_2$/WSe$_2$ moiré superlattices

*Hongyuan Li[1, 2, 3, 8], Shaowei Li[1, 3, 4, 8]\*, Mit H. Naik[1, 3], Jingxu Xie[1], Xinyu Li[1], Emma Regan[1, 2, 3], Danqing Wang[1, 2], Wenyu Zhao[1], Kentaro Yumigeta[5], Mark Blei[5], Takashi Taniguchi[6], Kenji Watanabe[7], Sefaattin Tongay[5], Alex Zettl[1, 3, 4], Steven G. Louie[1, 3], Michael F. Crommie[1, 3, 4]\*, Feng Wang[1, 3, 4]\**

[1]Department of Physics, University of California at Berkeley, Berkeley, CA, USA

[2]Graduate Group in Applied Science and Technology, University of California at Berkeley, Berkeley, CA, USA

[3]Materials Sciences Division, Lawrence Berkeley National Laboratory, Berkeley, CA, USA.

[4]Kavli Energy Nano Sciences Institute at the University of California Berkeley and the Lawrence Berkeley National Laboratory, Berkeley, CA, USA

[5]School for Engineering of Matter, Transport and Energy, Arizona State University, Tempe, AZ, USA

[6]International Center for Materials Nanoarchitectonics, National Institute for Materials Science, Tsukuba, Japan

[7]Research Center for Functional Materials, National Institute for Materials Science, Tsukuba, Japan

[8]These authors contributed equally: Hongyuan Li and Shaowei Li

1. **DFT calculation of conduction flat bands**

The near-0° twisted WS$_2$/WSe$_2$ moiré superlattice undergoes a structural reconstruction which leads to out-of-plane buckling of the two layers and in-plane strains which have been studied in a previous work[1]. Here, we study the influence of these reconstructions on the electronic structure of the conduction band states which is relevant to the experiment. The band structure (including spin-orbit coupling) of the reconstructed moiré pattern is plotted in the moire supercell BZ in Figure S1 (a). The three lowest-energy flat bands are states derived from the K point of the WS$_2$ layer and are strongly localized in the moiré pattern as shown in Fig. S1b, 1c and 1d, respectively. The discharging behavior observed in the experiment is that of an electron in the c1 band in which the wavefunction is localized on the WS$_2$ layer at the B$^{Se/W}$ site (as shown in Figure S1b).

For the simulation, the moiré superlattice is constructed using a 25x25 supercell of WSe$_2$ and a 26x26 supercell of WS$_2$ to ensure commensurability. The structural reconstruction is simulated using forcefield relaxation of the moiré superlattice as implemented in the LAMMPS[2] package. The intralayer interactions were modeled using the Stillinger-Weber[3, 4] forcefield and the interlayer interactions were studied using the Kolmogorov-Crespi[5, 6] forcefield. The force minimization was performed using the conjugate gradient method with tolerance of $10^{-6}$ eV/Å. The electronic structure of the reconstructed superlattice is studied using density functional theory[7] calculations as implemented in the Siesta[8] package. A supercell size of 21 Å was used in the out-of-plane direction to overcome spurious interaction between periodic images. A double-$\zeta$ plus polarization basis is used to expand the wavefunctions. Norm-conserving pseudopotentials[9] are used in the simulation and the exchange-correlation functional[10] is approximated using the local density approximation. Only the γ point was sampled in the supercell BZ to obtain the self-consistent charge density.

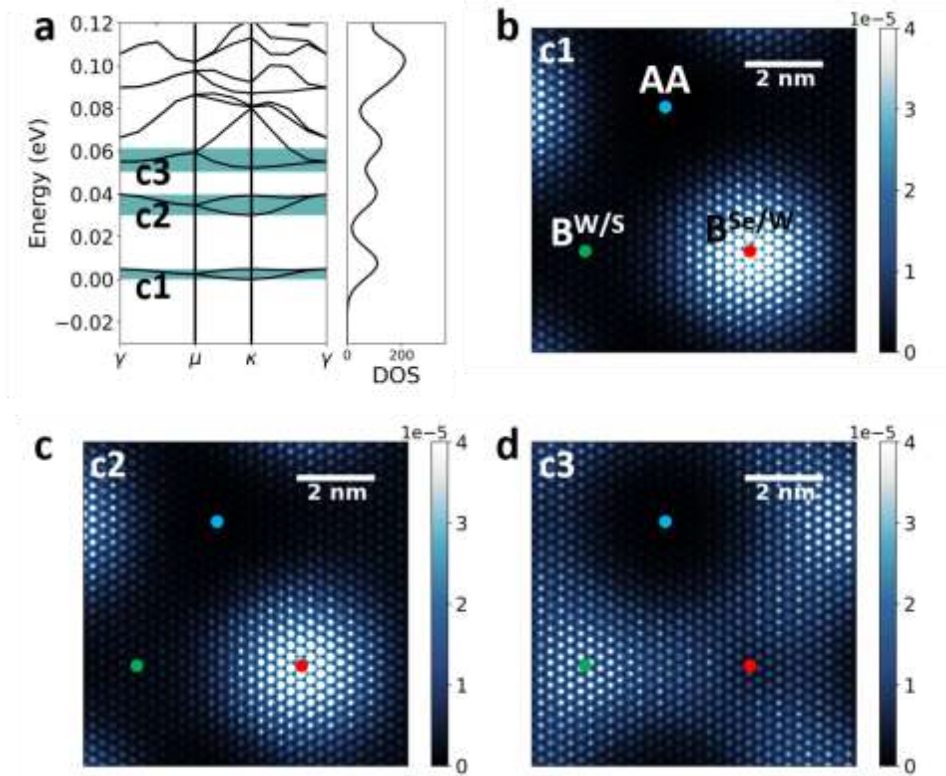

**Figure S1. DFT calculation of the $WS_2/WSe_2$ conduction flat bands**. **a.** DFT-calculated band structure for the conduction bands. The three lowest-energy flat bands are labeled c1, c2 and c3, respectively. **b-d.** Distribution of the local density of states for (**b**) c1, (**c**) c2, and (**d**) c3. Each map is integrated over the energy range indicated by the corresponding green stripe shown in **a**. The three stacking types (AA, $B^{Se/W}$, and $B^{W/S}$) are labeled in the maps. The lowest-energy flat band (c1) is localized at the $B^{Se/W}$ site.

## 2. Determination of the hopping energy, *t*, by fitting a tight binding model

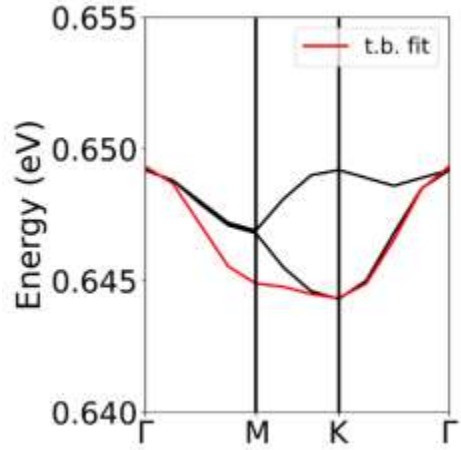

**Figure S2. Determination of the hopping energy t by fitting the a binding model.** Black line corresponds to the band structure of the lowest-energy conduction flat band calculated by DFT (c1 shown in Fig. S1a). The red line corresponds to a fitted band calculated by a tight-binding model on a triangular lattice, with the expression $E(\mathbf{k}) = \epsilon - 2t \cdot [\cos\left(\frac{a}{2}(k_x + \sqrt{3}k_y)\right) + \cos\left(\frac{a}{2}(k_x - \sqrt{3}k_y)\right) + \cos(ak_x)]$. The fitted hopping energy, t, is -0.555meV.

## 3. Extracting $\Delta V_b^D$ and $\Delta V_b^T$ via high-resolution dI/dV spectra mapping

We performed high-resolution dI/dV spectral mapping to measure the gaps between the cascade discharging peaks at high-symmetry D and T points ($\Delta V_b^D$ and $\Delta V_b^T$) as shown in Fig. S3. In order to more precisely obtain the discharging peak positions at the D and T points, we performed dI/dV spectroscopy measurements on a grid with spatial step size < 0.7 Å and bias step size < 14 mV. Figs. S3a-S3p show the evolution of the dI/dV maps sliced at different biases. Consistent with the results shown in the main text, three neighboring discharging rings expand and cross each other as the bias is increased. The nearest neighbor interactions induce the split cascade discharging peaks at the T and D points. The high spatial resolution allows us to more easily determine the T and D points (labeled in Fig. S3f and S3i). The corresponding spectra are shown in Fig. S3q (T point) and S3r (D point), with the discharging peaks labeled with blues arrows. Here the discharging peak width is ~40mV, much larger than the bias step size (14 mV).

Multiple moire sites are measured to obtain the statistical $\Delta V_b^D$ and $\Delta V_b^T$ data shown in the main text.

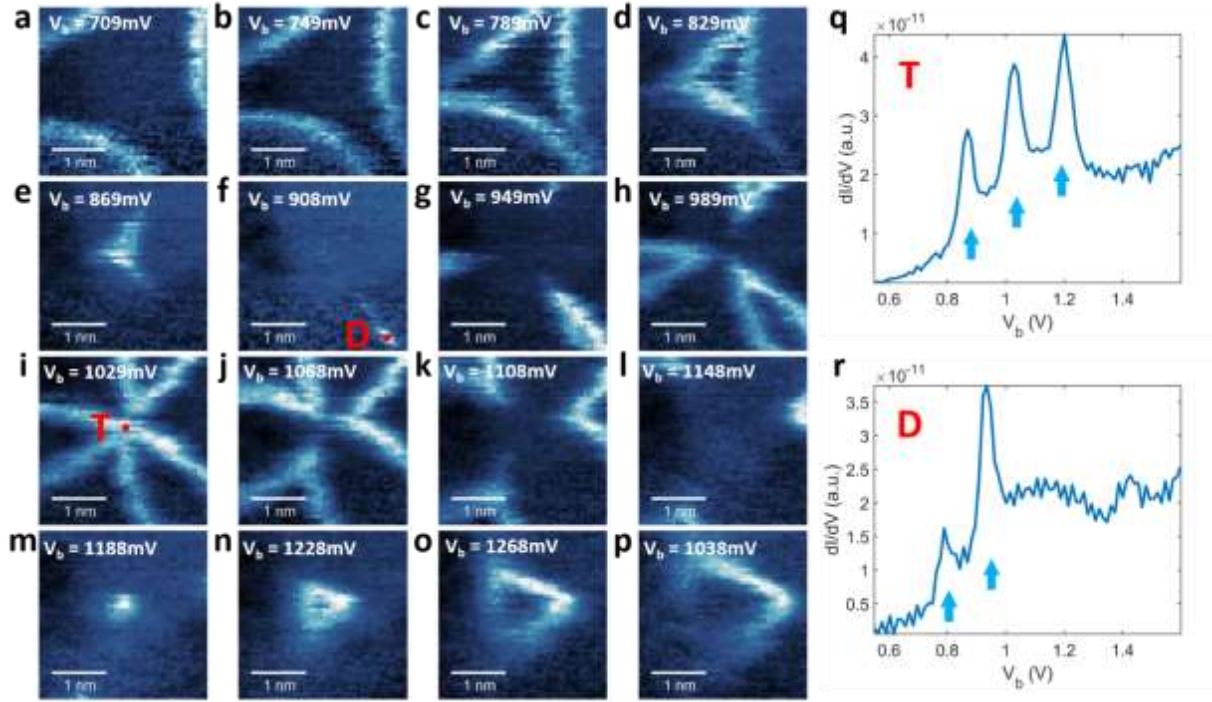

**Figure S3. High-resolution dI/dV spectra mapping around the T and D points**. **a-p**. Evolution of dI/dV mapping with increased bias from from 709mV to 1038mV. The positions of the two-ring crossing point (D) and three-ring crossing point (T) are labeled in **f** and **i**. $V_g = 45V$. The dI/dV spectroscopy mapping measurements were performed on a square grid. At each point, the tip-height was first fixed at the setpoint $V_b = -3.05V$ and $I = 104pA$, and then the dI/dV spectra at this point is measured under open-loop conditions. **q,r**. dI/dV spectra measured at the (**q**) T and (**r**) D points (pictured in **f** and **g**). The cascade discharging peaks are labeled with blue arrows.

## 4. Electrostatic simulation of $\alpha$ and $\beta$

Here we describe our determination of the values of $\alpha$ and $\beta$ through the use of electrostatic simulations. The system geometry of our model is illustrated in Fig. S4a. The STM tip is approximated as a metallic cone with half angle $\theta$. The apex-sample distance is H. The WS$_2$/WSe$_2$ heterostructure is approximated by two separate regions. The first region is a circular area with radius R$_{hole}$ surrounding the moire sites of interest (MSOI). The charge configuration in this region is set by the charging/discharging events described in the main text. The second region is the area outside the circle. The carrier doping in this region is relatively high and we can approximate it as a thin metal sheet. The Si back gate is regarded as an infinitely large metal plate and is separated from the TMD heterostructure plane by a distance d=320nm. The dielectric constants for the space above and below TMD heterostructure are assumed to be 1 (vaccum) and 4.2 (hBN and SiO$_2$).

As mentioned in the main text, the potential energy shift $v$ induced by V$_b$ and V$_g$ has the form

$$v = \alpha \cdot eV_b - \beta \cdot eV_g,$$

where we have neglected the moire site index i. We determine the values of $\alpha$ and $\beta$ via the following method: first, to determine $\alpha$ we ground the tip and the metallic plane and apply a 1V potential to the Si back gate. The value of the induced electric potential at the position of the moire electron (dot at the hole center) then equals $\alpha$. To determine $\beta$, we ground the Si back gate and the metallic plane and apply a 1V potential to the tip. The resulting value of the electric potential at the hole center then equals $\beta$.

Here we describe how to interpret the simulation results and compare them to the experimental data. Our simulation results should be fitted to the variation of the slopes of the dispersive discharging peak (for example Figs. 4d-g). As discussed in the main text, the slopes are equal to $\alpha(r_t)/\beta$. When the tip is close to the MSOI (for example at $r_t=0$) both H and $\theta$ significantly impact $\alpha$. However, when $r_t$ is large, the impact of the tip height, H, on $\alpha$ is negligible since H plays a negligible role in determining the distance between MSOI and the tip apex. Therefore, the shape of the curve $\alpha(r_t)/\beta$ vs $r_t$ can help us to choose the correct H and $\theta$ to match our experimental results. Finally, the radius of the hole $R_{hole}$ determines the screening strength of the surrounding TMD heterostructure on the external potential applied by the tip and the back gate. The value of $R_{hole}$ therefore does not significantly affect the shape of the curve $\alpha(r_t)/\beta$ vs $r_t$ (as long as $r_t$ is smaller than $R_{hole}$ (in order of magnitude as the moire period), a condition that is satisfied here).

Our electrostatic simulations were performed using COMSOL. The distribution of $\alpha(r_t)/\beta$ as the function of $r_t$ is well reproduced by using the following parameters: $\theta = 30°$, H=0.8 nm, and $R_{hole}$=8nm. Fig.S4b shows the simulated $\alpha$ (blue) and $\beta$ (orange) as functions of $r_t$. As expected, $\beta$ has a negligible change with the increase of $r_t$. The simulated ratio $\alpha(r_t)/\beta$ is plotted in Fig. S4c (orange curve) and is seen to compare well to the experimental data for $\alpha(r_t)/\beta$ (blue dots). The simulation yields $\alpha(r^T) \approx 0.18$, $\alpha(r^D) \approx 0.16$, and $\alpha(0) \approx 0.33$.

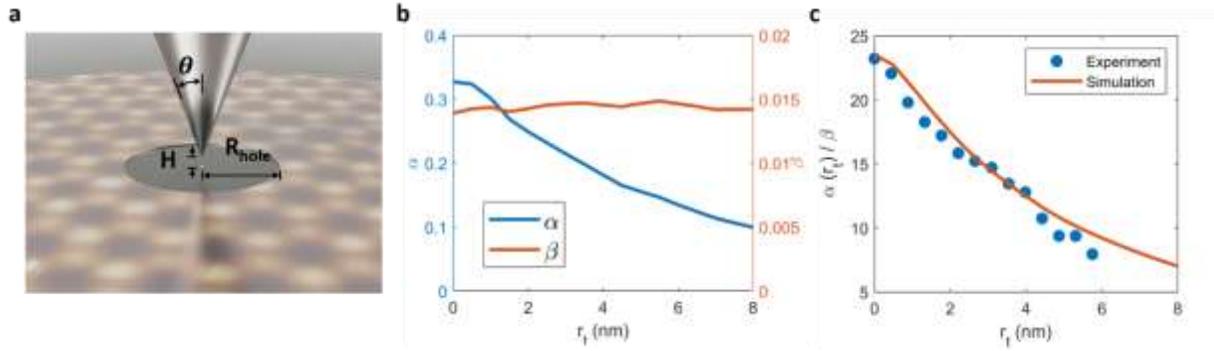

**Figure S4. Electrostatic simulation of $\alpha$ and $\beta$. a**. Electrostatic model of the tip-moire system. The tip apex is taken as a metallic cone with half angle $\theta$ and height H above the moiré site of interest (MSOI). The surrounding TMD heterostructure is assumed to be an infinitely large metallic plane with a hole of radius $R_{hole}$ around the MSOI (the dot located at the hole center). The Si back gate is regarded as an infinitely large metal plate and is separated from the TMD heterostructure plane by a distance d=320nm (not shown in **a**). **b**. Simulated $\alpha$ (blue) and $\beta$ (orange) as functions of $r_t$. As expected, $\beta$ is seen to be nearly independent of $r_t$ in our simulation range. **c**. Experimentally measured (blue points) and simulated (orange curve) values of $\alpha(r_t)/\beta$ as a function of $r_t$. The experiment and simulation show good agreement for the values H = 0.8 nm, $\theta = 30°$, and $R_{hole}$ = 8 nm.

**Reference:**

1. Li, H.; Li, S.; Naik, M. H.; Xie, J.; Li, X.; Wang, J.; Regan, E.; Wang, D.; Zhao, W.; Zhao, S. *Imaging moiré flat bands in three-dimensional reconstructed WSe 2/WS 2 superlattices. Nature Materials* **2021**, 1-6.
2. Plimpton, S. *Computational limits of classical molecular dynamics simulations. Computational Materials Science* **1995,** 4, (4), 361-364.
3. Stillinger, F. H.; Weber, T. A. *Computer simulation of local order in condensed phases of silicon. Physical review B* **1985,** 31, (8), 5262.

4.	Jiang, J.-W., *Handbook of Stillinger-Weber Potential Parameters for Two-Dimensional Atomic Crystals*. BoD–Books on Demand: 2017.
5.	Naik, M. H.; Maity, I.; Maiti, P. K.; Jain, M. *Kolmogorov–Crespi potential for multilayer transition-metal dichalcogenides: capturing structural transformations in moiré superlattices. The Journal of Physical Chemistry C* **2019,** 123, (15), 9770-9778.
6.	Kolmogorov, A. N.; Crespi, V. H. *Registry-dependent interlayer potential for graphitic systems. Physical Review B* **2005,** 71, (23), 235415.
7.	Kohn, W.; Sham, L. J. *Self-consistent equations including exchange and correlation effects. Physical review* **1965,** 140, (4A), A1133.
8.	Soler, J. M.; Artacho, E.; Gale, J. D.; García, A.; Junquera, J.; Ordejón, P.; Sánchez-Portal, D. *The SIESTA method for ab initio order-N materials simulation. Journal of Physics: Condensed Matter* **2002,** 14, (11), 2745.
9.	Troullier, N.; Martins, J. L. *Efficient pseudopotentials for plane-wave calculations. Physical review B* **1991,** 43, (3), 1993.
10.	Ceperley, D. M.; Alder, B. J. *Ground state of the electron gas by a stochastic method. Physical Review Letters* **1980,** 45, (7), 566.